\documentclass[12pt,preprint]{aastex}

\shorttitle{Determining Field Orientation of CMEs}

\shortauthors{Liu et al.}

\begin{document}

\title{Determining the Magnetic Field Orientation of Coronal Mass Ejections
from Faraday Rotation}

\author{Y. Liu\altaffilmark{1,2}, W. B. Manchester IV\altaffilmark{3},
J. C. Kasper\altaffilmark{1}, J. D. Richardson\altaffilmark{1,2}, and
J. W. Belcher\altaffilmark{1}}

\altaffiltext{1}{Kavli Institute for Astrophysics and Space Research,
Massachusetts Institute of Technology, Cambridge, MA 02139, USA;
liuxying@mit.edu.}

\altaffiltext{2}{State Key Laboratory of Space Weather, Chinese
Academy of Sciences, Beijing 100080, China.}

\altaffiltext{3}{Center for Space Environment Modeling, University of
Michigan, Ann Arbor, MI 48109, USA.}

\begin{abstract}
We describe a method to measure the magnetic field orientation of
coronal mass ejections (CMEs) using Faraday rotation (FR). Two basic
FR profiles, Gaussian-shaped with a single polarity or ``N"-like with
polarity reversals, are produced by a radio source occulted by a
moving flux rope depending on its orientation. These curves are
consistent with the Helios observations, providing evidence for the
flux-rope geometry of CMEs. Many background radio sources can map
CMEs in FR onto the sky. We demonstrate with a simple flux rope that
the magnetic field orientation and helicity of the flux rope can be
determined 2-3 days before it reaches Earth, which is of crucial
importance for space weather forecasting. An FR calculation based on
global magnetohydrodynamic (MHD) simulations of CMEs in a background
heliosphere shows that FR mapping can also resolve a CME geometry
curved back to the Sun. We discuss implementation of the method using
data from the Mileura Widefield Array (MWA).
\end{abstract}


\section{Introduction}

Coronal mass ejections (CMEs) are recognized as primary drivers of
interplanetary disturbances. The ejected materials are often
associated with large southward magnetic fields which can reconnect
with geomagnetic fields and produce storms in the terrestrial
environment \citep{dungey61, gosling91}. Determination of the CME
magnetic field orientation is thus of crucial importance for space
weather forecasting. However, nearly all atoms are ionized at the
coronal temperature $\sim 2\times 10^6$ K, making it difficult to
detect the coronal magnetic field through Zeeman splitting of
spectral lines as is routinely done for the photospheric field. A
typical way to estimate the coronal magnetic field above 2
$R_{\odot}$ ($R_{\odot}$ being the solar radius) is theoretical
extrapolation using the photospheric fields as boundary conditions,
which can only be checked by comparison to the field strength
measured from radio bursts and the orientation determined from soft
X-ray observations. The field orientation is also hard to infer from
white-light coronagraph images. Spacecraft near the first Lagrangian
point (L1) measure the local fields but can only give a warning time
for arrival at Earth of $\sim 30$ minutes \citep{vogt06, weimer02}.

A possible method to measure the coronal magnetic field is Faraday
rotation (FR), the rotation of the polarization plane of a radio wave
as it traverses a magnetized plasma. The first FR experiment was
conducted in 1968 by Pioneer 6 during its superior solar conjunction
\citep{levy69}. The observed FR curve features a ``W"-shaped profile
over a time period of 2-3 hours with rotation angles up to
40$^{\circ}$ from the quiescent baseline. This FR event was
interpreted as a coronal streamer stalk of angular size 1-2$^{\circ}$
\citep{woo97}, but \cite{patzold98} argue that the FR curve is
produced by the passage of a series of CMEs. Joint coronagraph
observations are needed to determine whether an FR transient is
caused by CMEs. Subsequent FR observations by the Pioneer and Helios
spacecraft reveal important information on the quiet coronal field
\citep{stelzried70, patzold87} and magnetic fluctuations
\citep{hollweg82, efimov96, andreev97, chashei99, chashei00}. FR
fluctuations are currently the only source of information for the
coronal field fluctuations. Independent knowledge of the electron
density, however, is needed in order to study the background field
and fluctuations.

Joint coronagraph and FR measurements of CMEs were also conducted
when the Helios spacecraft, with a downlink signal at a wavelength
$\lambda=13$ cm, was occulted by CME plasma. \cite{bird85} establish
a one-to-one correspondence between the SOLWIND white-light
transients and FR disturbances for 5 CMEs. Figure~1 displays the time
histories of FR and spectral broadening for two CMEs. Note that the
spectral broadening is proportional to the plasma density
fluctuations; the increased spectral broadening is consistent with
the enhanced density fluctuations within CMEs and their sheath
regions \citep{liu06b}. The FR through the 23 October 1979 CME shows
a curve (note a data gap) which seems not to change sign during the
CME passage; a single sign in FR indicates a monopolar magnetic
field. The 24 October 1979 CME displays an FR curve which is roughly
``N"-like across the zero rotation angle, indicative of a dipolar
field. Other CMEs in the work of \cite{bird85} give similar FR
curves, either an ``N"-type or a waved shape around the zero level.
Based on a simple slab model for CMEs, the mean transient field
magnitude is estimated to be 10 - 100 mG scaled to $2.5 R_{\odot}$,
which seems larger than the mean background field. The CME field
geometry, as implied by these FR curves, will be discussed below.
These features demonstrate why radio occultation measurements are
effective in detecting CMEs.

FR experiments using natural radio sources, such as pulsars and
quasars, have also been performed. FR observations of this class were
first conducted by \cite{bird80} during the solar occultation of a
pulsar. The advantage of using natural radio sources is that many of
these sources are present in the vicinity of the Sun and provide
multiple lines of sight which can be simultaneously probed by a radio
array. We can thus make a two-dimensional (2-D) mapping of the solar
corona and the inner heliosphere with an extended distribution of
background radio sources.

In this paper, we show a method to determine the magnetic field
orientation of CMEs using FR. This method enables us to acquire the
field orientation 2 - 3 days before CMEs reach Earth, which will
greatly improve our ability to forecast space weather. The data
needed to implement this technique will be available from the Mileura
Widefield Array (MWA) \citep{mwa05}. The magnetic structure obtained
from MWA measurements with this method will fill the missing link in
coronal observations of the CME magnetic field and also place strong
constraints on CME initiation theories.

\section{Modeling the Helios Observations}

The FR technique uses the fact that a linearly polarized radio wave
propagating through a magnetized plasma will undergo a rotation in
its plane of polarization. The rotation angle is given by $\Omega =
\lambda^2RM$, where $\lambda$ is the wavelength of the radio wave.
The rotation measure, $RM$, is expressed as
\begin{equation}
RM = \frac{e^3}{8\pi^2\epsilon_0m_e^2c^3} \int n_e \textbf{B} \cdot
d\textbf{s},
\end{equation}
where $e$ is the electron charge, $\epsilon_0$ is the permittivity of
free space, $m_e$ is the electron mass, $c$ is the speed of light,
$n_e$ is the electron density, $\textbf{B}$ is the magnetic field,
and $d\textbf{s}$ is the vector incremental path defined to be
positive toward the observer. FR responds to the magnetic field,
making it a useful tool to probe the coronal transient and quiet
magnetic fields. Note that the polarization vector may undergo
several rotations across the coronal plasma. Measurements at several
frequencies are needed to break the degeneracy; observations as a
function of time can also help to trace the rotation through its
cycles.

In situ observations of CMEs from interplanetary space indicate that
CMEs are often threaded by magnetic fields in the form of a helical
flux rope \citep{burlaga81, burlaga88, lepping90}. This helical
structure either exists before the eruption \citep{chen96, kumar96,
gibson98, lin00}, as needed for supporting prominence material, or is
produced by magnetic reconnection during the eruption
\citep[e.g.,][]{mikic94}. The flux rope configuration reproduces the
white-light appearance of CMEs \citep{chen96, gibson98}. This
well-organized structure will display a specific FR signature easily
discernible from the ambient medium, but direct proof of the
flux-rope geometry of CMEs at the Sun has been lacking.

\subsection{Force-Free Flux Ropes}

Here we model the Helios observations using a cylindrically symmetric
force-free flux rope \citep{lun50} with
\begin{equation}
\textbf{B} = B_0J_0(\alpha r)\hat{z} + B_0HJ_1(\alpha r)\hat{\phi}
\end{equation}
in axis-centered cylindrical coordinates $(\hat{r}, \hat{\phi},
\hat{z})$ in terms of the zeroth and first order Bessel functions
$J_0$ and $J_1$ respectively, where $B_0$ is the field magnitude at
the rope axis, $r$ is the radial distance from the axis, and $H$
specifies the left-handed (-1) or right-handed (+1) helicity. We take
$\alpha r_0 = 2.405$, the first root of the $J_0$ function, so
$\alpha$ determines the scale of the flux-rope radius $r_0$. The
electron density is obtained by assuming a plasma beta $\beta=0.1$
and temperature $T = 10^5$ K, as implied by the extrapolation of in
situ measurements \citep[e.g.,][]{liu05, liu06a}. Combining
equations~(1) and (2) with a radio wave path gives the FR.

For simplicity, we consider a frame with the $x$-$y$ plane aligned
with the flux-rope cross section at its center and the $z$ axis along
the axial field. Figure~2 shows the diagram of the flux rope with the
projected line of sight. The flux rope, initially at 4$R_{\odot}$
away from the Sun with a constant radius $r_0=3.6R_{\odot}$ and
length 20$R_{\odot}$, moves with a speed $v=500$ km s$^{-1}$ in the
$x$ direction across a radio ray path. The radio signal path makes an
angle $\theta$ with respect to the plane and $\phi$ with the motion
direction when projected onto the plane. The magnetic field strength
at the rope axis is adopted to be $B_0=25$ mG, well within the range
estimated from the Helios observations \citep{bird85}.

The resulting FR curves are displayed in Figure~3. A radio source
occulted by the moving flux rope gives two basic types of FR curves,
Gaussian-shaped and ``N"-shaped (or inverted ``N") depending on the
orientation of the radio wave path with respect to the flux rope.
When the radio signal path is roughly along the flux rope (say, for
$\phi=45^{\circ}$ and $\theta=60^{\circ}$ as shown in the right
panel), the axial field overwhelms the azimuthal field along the
signal path, so the FR curve would be Gaussian-like, indicative of a
monopolar field. For a signal path generally perpendicular to the
flux rope, the azimuthal field dominates and changes sign along the
path, so the rotation curve would be ``N" or inverted ``N" shaped
with a sign change (left panel), suggestive of a dipolar field. These
basic curves are consistent with the Helios measurements. Two
adjacent flux ropes with evolving fields could yield a ``W"-shaped
curve as observed by Pioneer 6 \citep{levy69, patzold98}. The time
scale and magnitude of the observed FR curves are also reproduced.
When $\theta=0^{\circ}$, the line of sight is within the plane.
Varying $\phi$ gives a variety of time scales of FR, ranging from
$\sim 3$ to more than 10 hours, but the peak value of FR is fixed at
$\sim 57^{\circ}$. These numbers are consistent with the Helios data
shown in the right panel of Figure~1. When $\theta$ is close to
$90^{\circ}$, the observer would be looking along the flux rope. The
axial field produces a strong FR, but decreasing $\theta$ will
diminish the rotation angle and make the curve more and more
``N"-like. The time scale, however, remains at 4 hours. For
$\phi=45^{\circ}$ and $\theta=40^{\circ}$, the rotation angle is up
to $140^{\circ}$, in agreement with the Helios data shown in the left
panel of Figure~1.

\subsection{Non-Force-Free Flux Ropes}

A non-force-free flux rope could give more flexibility in the field
configuration. Consider a magnetic field that is uniform in the $z$
direction in terms of rectangular coordinates. Since $\nabla\cdot
\textbf{B}= 0$, the magnetic field can be expressed as
\begin{equation}
{\bf B} = \left({\partial A \over \partial y}, -{\partial A \over
\partial x}, B_z\right),
\end{equation}
where the vector potential is defined as ${\bf A} = A(x, y) \hat{\bf
z}$. The MHD equilibrium, ${\bf j}\times {\bf B} - \nabla p = 0$,
gives \citep[e.g.,][]{sturrock94}
\begin{equation}
{\partial^2A \over \partial x^2} + {\partial^2A \over \partial y^2} =
-\mu_0{d \over dA}\left(p + {B_z^2 \over 2\mu_0}\right) = -\mu_0j_z,
\end{equation}
where $\mu_0$ is the permeability of free space, $p$ is the plasma
thermal pressure, and $j_z$ is the $z$ component of the current
density. Equation~(4) is known as the Grad-Shafranov equation. We see
from this equation that $p$, $B_z$ and hence $j_z$ are a function of
$A$ alone. A special form of this equation $\nabla^2 \tilde{A} =
\exp(-2\tilde{A})$ (in properly scaled units) has the solution
\citep[e.g.,][]{schindler73}
\begin{equation}
\tilde{A} = \ln\left[\alpha\cos \tilde{x} + \sqrt{1+\alpha^2}\cosh
\tilde{y}\right].
\end{equation}
This nonlinear solution has been called the periodic pinch since it
has the form of a 2-D neutral sheet perturbed by a periodic chain of
magnetic islands centered in the current sheet. Here $\tilde{A}$,
$\tilde{x}$ and $\tilde{y}$ are dimensionless quantities, and
$\alpha$ is a free parameter that can be used to control the aspect
ratio of the magnetic islands.

From equations~(3)-(5) we obtain
\begin{equation}
j_z = - \frac{B_0}{\mu_0 L_0}\exp\left(\frac{-2A}{B_0L_0}\right),
\end{equation}
\begin{equation}
B_x = B_0\frac{\sqrt{1+\alpha^2} \sinh(y/L_0)}{\alpha\cos(x/L_0) +
\sqrt{1+\alpha^2}\cosh(y/L_0)},
\end{equation}
\begin{equation}
B_y = B_0\frac{\alpha\sin(x/L_0)}{\alpha\cos(x/L_0) +
\sqrt{1+\alpha^2}\cosh(y/L_0)},
\end{equation}
where $B_0$ and $L_0$ are scales of the field magnitude and length,
respectively. The axial field $B_z$ and the thermal pressure can be
obtained from $\frac{d}{dA}(p + \frac{B_z^2}{2\mu_0}) = j_z$, which
gives $$p + \frac{B_z^2}{2\mu_0} =
\frac{B_0^2}{2\mu_0}\exp\left(\frac{-2A}{B_0L_0}\right) +
\frac{B_1^2}{2\mu_0},$$ where $B_1$ is an arbitrary constant.
Assuming a factor $\varepsilon$ in the partition of the total
pressure, we have
\begin{equation}
p =
\varepsilon\frac{B_0^2}{2\mu_0}\left[\left(\alpha\cos\frac{x}{L_0} +
\sqrt{1+\alpha^2}\cosh\frac{y}{L_0}\right)^{-2} + \frac{B_1^2}{B_0^2}
\right],
\end{equation}
\begin{equation}
B_z = \pm\sqrt{1-\varepsilon}B_0\left[\left(\alpha\cos\frac{x}{L_0} +
\sqrt{1+\alpha^2}\cosh\frac{y}{L_0}\right)^{-2} + \frac{B_1^2}{B_0^2}
\right]^{1/2}.
\end{equation}
Adjusting the parameters $\alpha$ and $\varepsilon$ gives a variety
of flux rope configurations, circular and non-circular, force-free
and non-force-free.

A flux rope of this kind is displayed in Figure~4. As can be seen,
this flux rope lies within a current sheet. To single out the flux
rope, we require $0\leq x \leq 2\pi L_0$ and $-\pi L_0/2\leq y \leq
\pi L_0/2$ initially, where $L_0 = 1.5R_{\odot}$. The flux rope is
still $20R_{\odot}$ long, moving with $v = 500$ km s$^{-1}$ across
the line of sight. Other parameters are assumed to be $B_0 = 10$ mG,
$B_1 = 0$, $\alpha=2$, $\varepsilon=0.1$, and the temperature $T =
10^5$ K. Figure~5 shows the calculated FR. These curves are generally
similar to those for a cylindrically symmetric force-free flux rope.
Unlike the force-free flux-rope counterpart, the FR curves show a
smooth transition from the zero angle to peak values. In addition,
they are narrower in width, which may result from fields and
densities which are more concentrated close to the axis. Note that
the field magnitude is $\sim 40$ mG at the axis of the non-force-free
flux rope. These profiles can also qualitatively explain the Helios
observations.

The above results suggest that CMEs at the Sun manifest as flux
ropes, confirming what previously could only be inferred from in situ
data \citep{burlaga88, lepping90}.  They also reinforce the
connection of CMEs observed by coronagraphs with magnetic clouds
identified from in situ measurements.

\section{2-D Mapping of CMEs}

As demonstrated above, even a single radio signal path can give hints
on the magnetic structure of CMEs. Ambiguities in the flux rope
orientation cannot be removed based on only one radio ray path. The
power of the FR technique lies in having multiple radio sources,
especially when a 2-D mapping of CMEs onto the sky is possible.

\subsection{A Single Flux Rope}

For a flux-rope configuration, the magnetic field is azimuthal close
to the rope edge and purely axial at the axis. The rotation measure
would be positive through the part of the rope with fields coming
toward an observer and negative through the part with fields leaving
the observer, so the azimuthal field orientation can be easily
recognized with data from multiple lines of sight (radio ray paths).
A key role is played by the axial component, which tells us the
helicity of the flux rope. Consider a force-free flux rope for
simplicity. For points on a line parallel to the rope axis within the
flux rope, the field direction as well as the magnitude is the same.
The fields on this line would make different angles with a variety of
radio signal paths since the signal path is always toward the
observer. As long as the axial field component is strong enough,
these different angles will lead to a gradient in the rotation
measure along the rope.

Assuming an observer sitting at Earth, we calculate the FR pattern
projected onto the sky for a force-free flux rope viewed from many
radio sources. A flux rope has two possibilities for the axial field
direction, with each one accompanied by either a left-handed or
right-handed helicity. Plotted in Figure~6 are the four possible
configurations as well as their rotation measure patterns. The angle
$\theta_y$ defines the azimuthal angle of a line of sight with
respect to the Sun-Earth (observer) direction in the solar ecliptic
plane, while $\theta_z$ is the elevation angle of the line of sight
with respect to the ecliptic plane. The flux rope, with axis in the
ecliptic plane and perpendicular to the Sun-Earth direction, is
centered at 10$R_{\odot}$ from the Sun and has a radius of $r_0 =
8R_{\odot}$ and length $50R_{\odot}$. The magnetic field magnitude is
assumed to be 10 mG at the rope axis. The gradient effect in the
rotation measure along the flux rope is apparent in Figure~6 and it
produces a one-to-one correspondence between the flux-rope
configuration and the rotation measure pattern. The four
configurations of a flux rope can thus be uniquely determined from
the global behavior of the rotation measure, which gives the axial
field orientation and the helicity. In order to fully resolve the
flux rope, we have assumed $\sim 80$ radio sources per square degree
on the sky, but in practice a resolution of 250 times lower can give
enough information for the field orientation and helicity (see
Figure~7).

The FR mapping obtained from multiple radio sources can also help to
determine the speed and orientation of CMEs as they move away from
the Sun. This mapping is similar to coronagraph observations. While
the polarized brightness (Thomson-scattered, polarized component of
the coronal brightness) is sensitive to the electron density, FR
reacts to the magnetic field as well as the electron density and thus
may be able to track CMEs to a larger distance than white light
imaging. Figure~7 gives snapshots at different times of a tilted flux
rope moving outward from the Sun. A Sun-centered coordinate system is
defined such that the $x$ axis extends from the Sun to Earth, the $z$
axis is normal to and northward from the solar ecliptic plane, and
the $y$ axis lies in the ecliptic plane and completes the right
handed set. A force-free flux rope, initially centered at (2, 2,
2)$R_{\odot}$ in this frame and oriented at 30$^{\circ}$ from the
ecliptic plane and 70$^{\circ}$ from the Sun-Earth line, moves at a
speed 500 km s$^{-1}$ from the Sun along the direction with elevation
angle 10$^{\circ}$ and azimuthal angle 20$^{\circ}$. The flux rope
evolution is constructed by assuming a power law dependence with
distance $R$ (in units of AU) for the rope size and physical
parameters, i.e., $$r_0 = 0.2\times R^{0.78}~{\rm AU}$$ for the rope
radius, $$B_0 = 15\times R^{-1.5}~{\rm nT}$$ for the field magnitude
at the axis, and $$T = 3\times 10^4\times R^{-0.72}~{\rm K}$$ for the
temperature. The rope length is kept at 3 times the rope diameter,
and the plasma $\beta$ is kept at 0.1. Similar power-law dependences
have been identified by a statistical study of CME evolution in the
solar wind \citep{liu05, liu06a}, but note that the transverse size
of the flux-rope cross section could be much larger than the radial
width \citep{liu06c}.

The 2-D mapping has a pixel size of about 3.2 degrees. Even at such a
low resolution, the flux rope can be recognized several hours after
appearance at the Sun. The orientation of the flux rope with respect
to the ecliptic plane is apparent in the first few snapshots, but
note that this elevation angle may be falsified by the projection
effect. The gradient effect in the rotation measure along the flux
rope is discernable at 10 hours and becomes clearer around 20 hours.
A right-handed helicity with axial fields skewed upward can be
obtained from this gradient after a comparison with Figure~6 (top
left). When the flux rope is closer to Earth, its appearance
projected onto the sky becomes more and more deformed. Finally, when
Earth is within the flux rope (around 80 hours), an observer would
see two spots with opposite polarities produced by the ends of the
flux rope.

Note that the above conclusions are not restricted to cylindrically
symmetric force-free flux ropes. We have also used the non-force-free
solutions of the steady state Vlasov-Maxwell equations (see \S 2.2),
which unambiguously give the same picture. The FR technique takes
advantage of an axial magnetic field coupled with the azimuthal
component, which is the general geometry of a flux rope. This robust
feature makes possible a precise determination of the CME field
orientation. A curved flux rope with turbulent fields, however, may
need caution in determining the axial field direction (see below).

\subsection{MHD Simulations with Background Heliosphere}

The above FR calculation does not take into account the background
heliosphere. In this sense, the 2-D mapping may be considered as a
difference imaging between the transient and background heliospheres.
Here we use for the FR calculation 3-D ideal MHD simulations of a CME
propagating into a background heliosphere \citep{manchester04}. The
simulations are performed using the so-called Block Adaptive Tree
Solar-Wind Roe Upwind Scheme (BATS-R-US). A specific heating function
is assumed to produce a global steady-state model of the corona that
has high-latitude coronal holes (where fast winds come from) and a
helmet streamer with a current sheet at the equator. A twisted flux
rope with both ends anchored in the photosphere is then inserted into
the helmet streamer. Removal of some plasma in the flux rope
destabilizes the flux rope and launches a CME. The numerical
simulation with adaptive mesh refinement captures the CME evolution
from the solar corona to Earth. A 3-D view of the flux rope resulting
from the simulations is displayed in Figure~8. The magnetic field, as
represented by colored solid lines extending from the Sun, winds to
form a helical structure within the simulated CME. The field has a
strong toroidal (axial) component close to the axis but is nearly
poloidal (azimuthal) at the surface of the rope.

A fundamental problem in CME studies which remains to be resolved is
whether CMEs are magnetically connected to the Sun as they propagate
through interplanetary medium. Most theoretical modeling assumes a
twisted flux rope with two ends anchored to the Sun \citep{chen96,
kumar96, gibson98}. This scenario is suggested by energetic particles
of solar origin observed within a magnetic cloud \citep{kahler91}. An
isolated plasmoid is also a possible structure for CMEs
\citep{vandas93a, vandas93b}. The FR mapping is capable of removing
this ambiguity in that it can easily capture a flux-rope geometry
bent toward the Sun. To show this capability, we calculate the FR
mapping of the simulated CME in a background heliosphere. The MHD
model gives a time series of data cubes of $300 R_{\odot}$ in length.
We subtract the background from the rotation measure of the CME data
to avoid possible effects brought about by the finite domain.
Figure~9 shows the difference mapping of the rotation measure at a
resolution of $\sim 3.2$ degrees when the CME propagates a day ($\sim
70 R_{\odot}$) away from the Sun. The simulation data are rotated
such that the observer (projected onto the origin) can see the flux
rope curved to the Sun. The coordinates, $\theta_y$ and $\theta_z$,
are defined with respect to the observer. A flux rope extending back
to the Sun is apparent in the difference image. The outer arc with
positive rotation measures is formed by the azimuthal magnetic field
pointing to the observer while the inner arc with negative rotation
measures originates from the field with the opposite polarity. The
rotation measure difference is positive near the Sun, which is due to
a pre-existing negative rotation measure that becomes less negative
after the CME eruption.

A closer look at the image would also reveal asymmetric legs of the
flux rope. This effect, indicative of a right-handed helicity, is
created by the different view angles as described above. The nose of
the flux rope does not show a clear gradient in the rotation measure
because the view angles of this part are similar. In the case of the
two legs directed to the observer, two spots with contrary magnetic
polarities will be seen, so the curved geometry may also help to
clarify the field helicity.

\section{Summary and Discussion}

We have presented a method to determine the magnetic field
orientation of CMEs based on FR. Our FR calculations, either with a
simple flux rope or global MHD modeling, demonstrate the exciting
result that the CME field orientation can be obtained 2-3 days before
CMEs arrive at Earth, substantially longer than the warning time
achieved by local spacecraft measurements at L1.

The FR curves through the CME plasma observed by Helios can be
reproduced by a flux rope moving across a radio signal path. Two
basic FR profiles, Gaussian-shaped with a single polarity or
``N"-like with polarity reversals, indicate the orientation of the
flux rope with respect to the signal path. Force-free and
non-force-free flux ropes generally give the same picture, except
some trivial differences reflecting the field and density
distributions within a flux rope. The FR calculation with a radio
signal path, combined with the Helios observations, shows that CMEs
at the Sun appear as flux ropes.

2-D FR mapping of a flux rope using many radio sources gives the
field orientation as well as the helicity. The orientation of
azimuthal fields can be readily obtained since they yield rotation
measures with opposite polarities. The axial component of the
magnetic field creates a gradient in rotation measure along the flux
rope, with which the flux rope configurations can be disentangled.
Time-dependent FR mapping is also calculated for a tilted flux rope
propagating away from the Sun. The orientation of the flux rope as a
whole and its projected speed onto the sky can be determined from the
snapshots of the flux rope mapped in FR. We further compute the FR
mapping for a curved flux rope moving into a background heliosphere
obtained from 3-D ideal MHD simulations. It is shown that the FR
mapping can resolve a CME curved back to the Sun in addition to the
field orientation. Difference imaging is needed to remove the FR
contribution from the background medium.

The global FR map is a new technique for measuring the CME magnetic
field. This method can determine the magnetic field orientation of
CMEs without knowledge of the electron density. The electron density
could be inferred from Thomson scattering measurements made by the
SECCHI instrument (suite of wide angle coronagraphs) on STEREO which
has stereoscopic fields of view \citep{howard00}. With the joint
measurements of the electron density, the magnetic field strength can
be estimated.

Note that the above results are a first-order attempt to predict what
may be seen in FR. An actual CME likely shows a turbulent behavior
and may have multiple structures along the line of sight; the
rotation measure, an integral quantity along the line of sight, could
display similar signatures for different structures. Therefore,
interpretation of the FR measurements will be more complex than
suggested here. However, having an instantaneous, global map of the
rotation measure that evolves in time will be vastly superior to a
time profile along a single line of sight, and comparison with
coronagraph observations and actual measures of geoeffectiveness
(e.g., the $D_{st}$ index) for a series of real events will
eventually lead to the predictive capability proposed in this paper.

The present results also pave the way for interpreting future FR
observations of CMEs by large radio arrays, particularly those
operating at low frequencies \citep{lofar04, mwa05}. The MWA - Low
Frequency Demonstrator, specially designed for this purpose at 80-300
MHz, will feature wide fields of view, high sensitivity and
multi-beaming capabilities \citep{mwa05}. This array will be
installed in Western Australia (26.4$^{\circ}$S, 117.3$^{\circ}$E), a
radio quiet region. It will spread out $\sim 1.5$ km in diameter,
achieving $\sim 8000$ m$^2$ of collecting area at 150 MHz and a field
of view from 15$^{\circ}$ at 300 MHz to 50$^{\circ}$ at 80 MHz. The
point source sensitivity will be about 20 mJy for an integration time
of 1 s. The array is expected to monitor $\sim 300$ background radio
sources within 13$^\circ$ elongation ($\sim 50 R_{\odot}$) from the
Sun, providing a sufficient spatial sampling of the inner
heliosphere. In addition, this array will be able to capture a
rotation measure of $\sim 10^{-2}$ rad m$^{-2}$ and thus is
remarkably sensitive to the magnetic field. Science operations of the
array will start in 2009. Implementation of our method by such an
array would imply a coming era when the impact of the solar storm on
Earth can be predicted with small ambiguities. It could also fill the
missing link in coronal observations of the CME magnetic field, thus
providing strong constraints on CME initiation theories.

\acknowledgments
The research was supported by NASA contract 959203
from JPL to MIT and NASA grant NAG5-11623. This work was also
supported by the CAS International Partnership Program for Creative
Research Teams.

\clearpage

\begin{figure}
\centerline{\includegraphics[width=17pc, height=20pc]{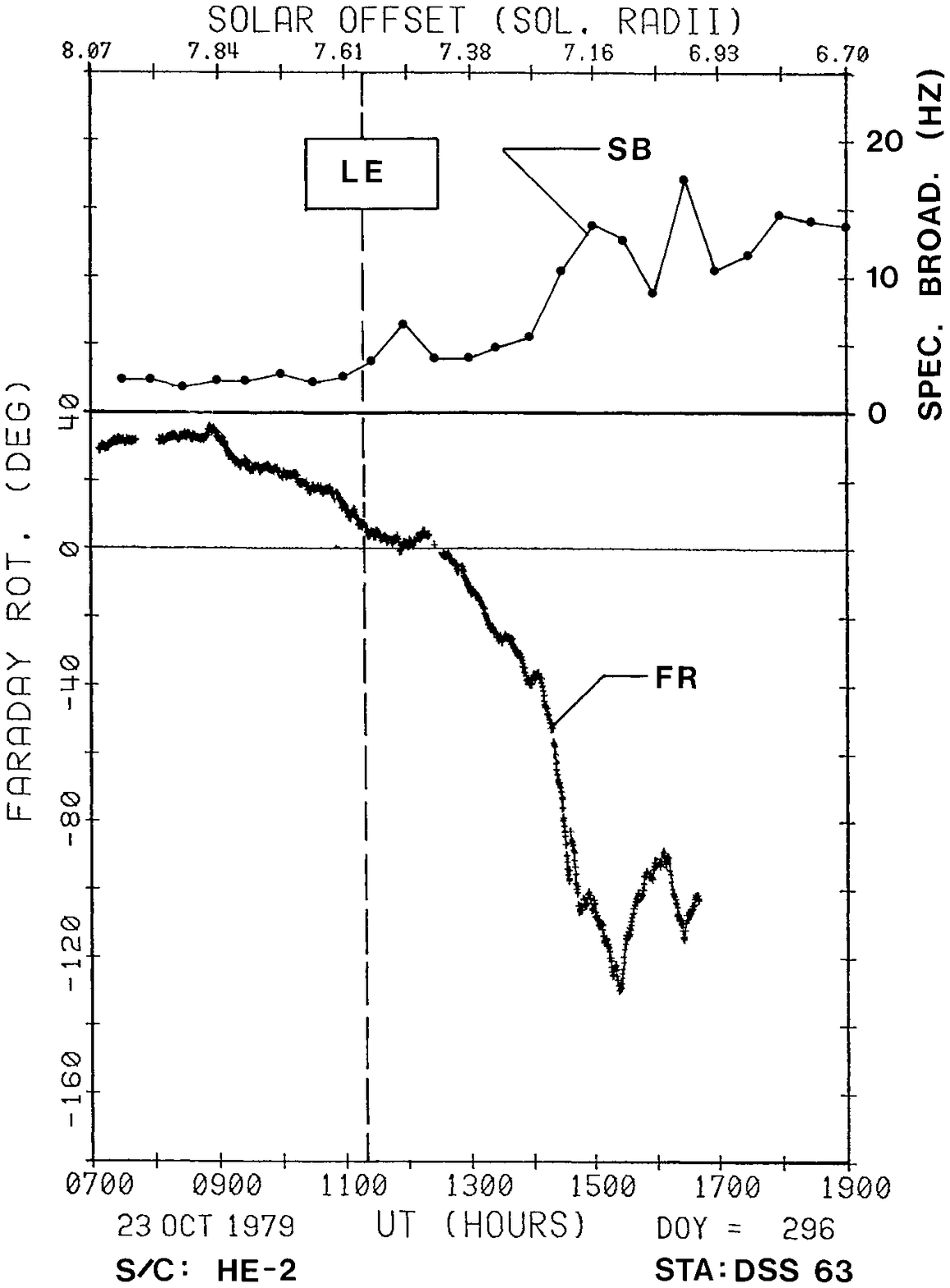}
\includegraphics[width=16pc, height=20pc, angle=-0.5]{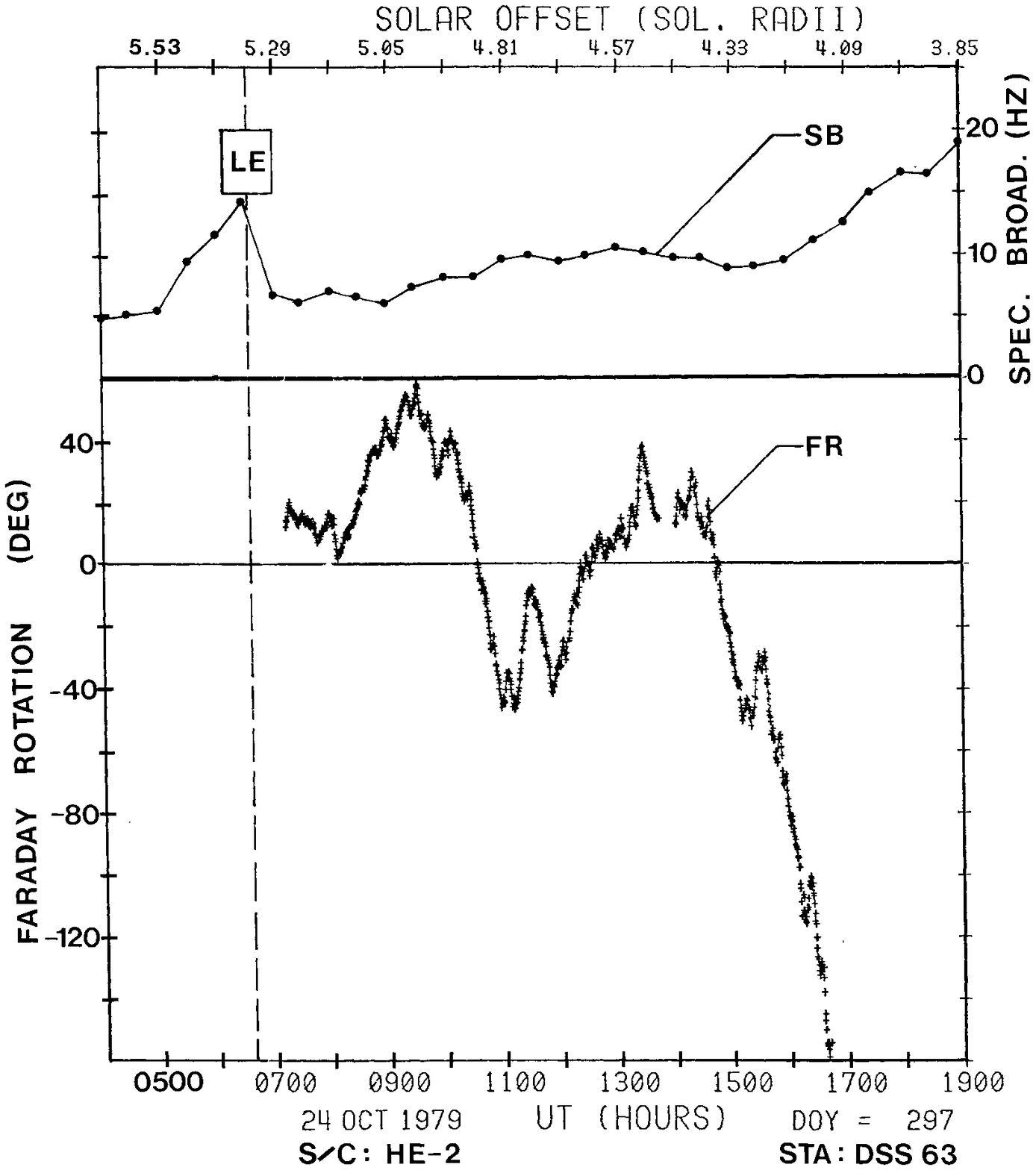}}
\caption{Time profiles of FR (bottom) and spectral broadening (top)
of the Helios 2 signal during the CMEs of 23 October 1979 (left) and
24 October 1979 (right) recorded at the Madrid station DSS 63. The
apparent solar offset of Helios 2 is given at the top. The dashed
vertical line indicates the arrival time of the CME leading edge with
uncertainties given by the width of the box ``LE''. Large deviations
in FR following the leading edge indicate the arrival of the CME's
bright core. Reproduced from \cite{bird85}.}
\end{figure}

\clearpage

\begin{figure}
\plotone{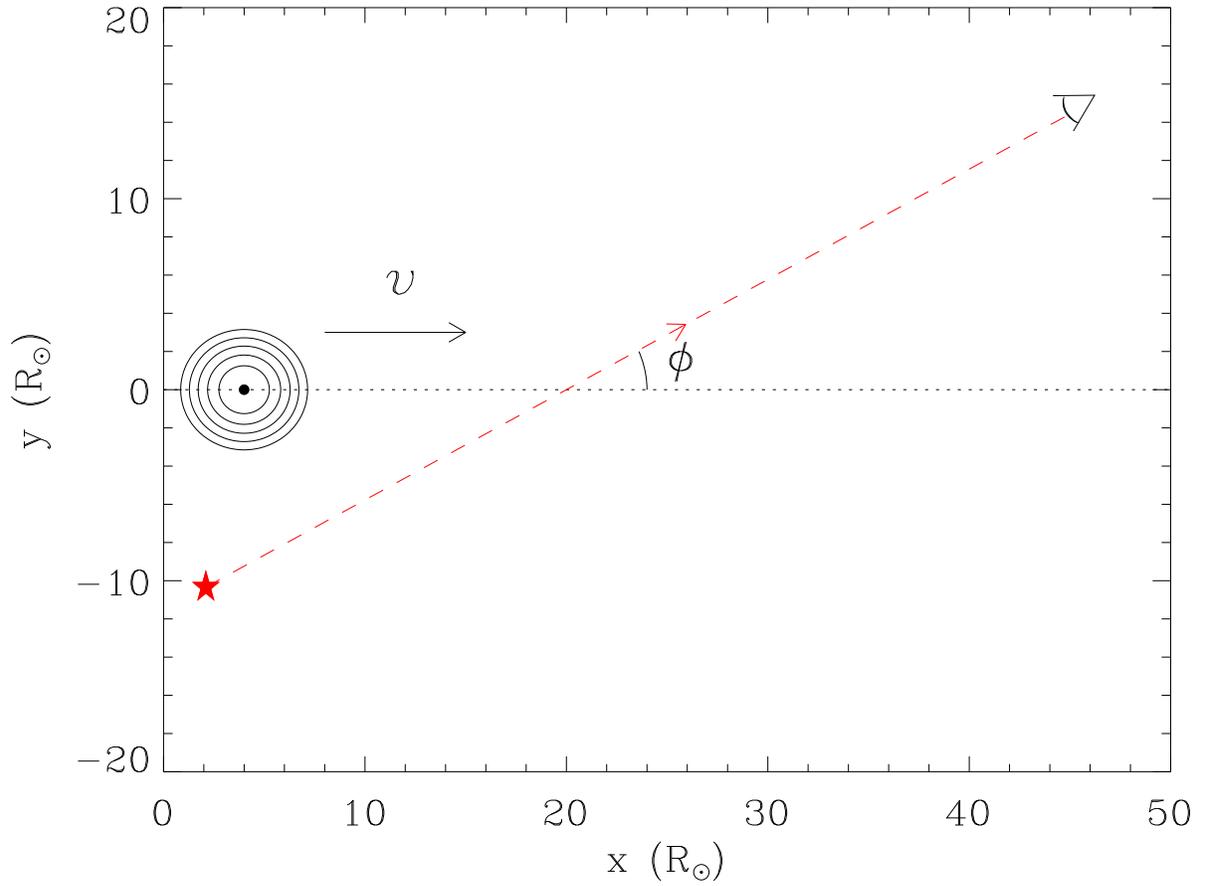}
\caption{Schematic diagram of a force-free flux rope
and the line of sight from a radio source to an observer projected
onto the plane of the flux-rope cross section. The flux rope moves at
a speed $v$ across the line of sight which makes an angle $\phi$ with
the motion direction.}
\end{figure}

\clearpage

\begin{figure}
\plotone{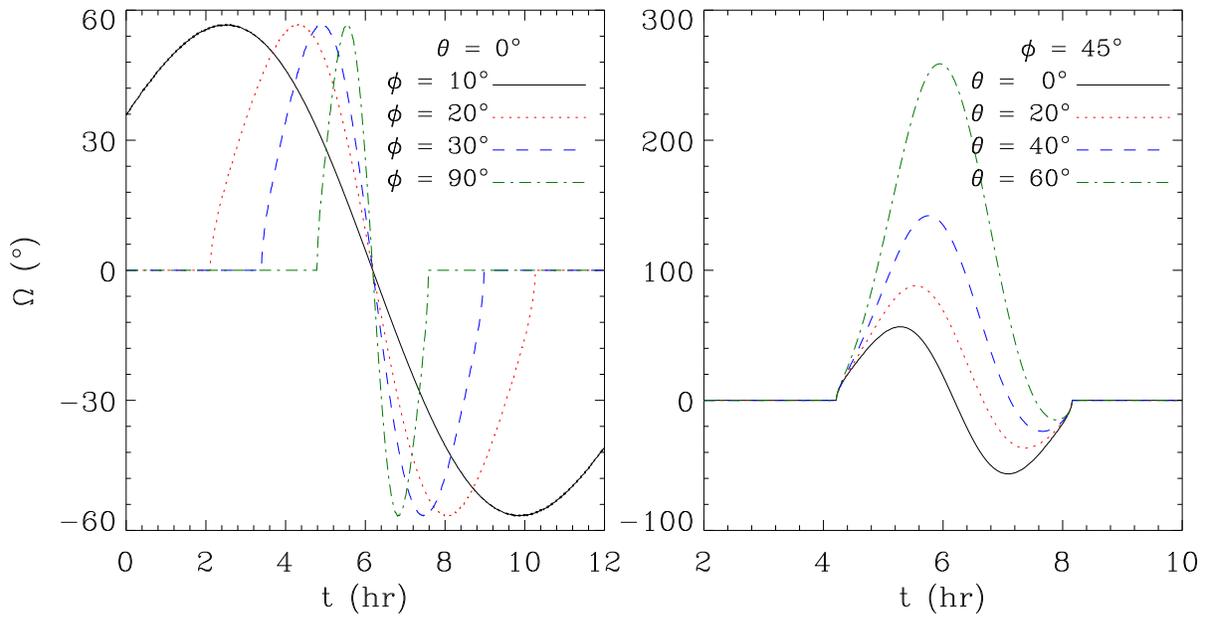}
\caption{FR at $\lambda=13$ cm through the
force-free flux rope as a function of time. Left is the rotation
angle with $\theta$ fixed to 0$^{\circ}$ and $\phi=[10^{\circ},
20^{\circ}, 30^{\circ}, 90^{\circ}$], and right is the rotation angle
with $\phi$ fixed to 45$^{\circ}$ and $\theta=[0^{\circ}, 20^{\circ},
40^{\circ}, 60^{\circ}$].}
\end{figure}

\clearpage

\begin{figure}
\plotone{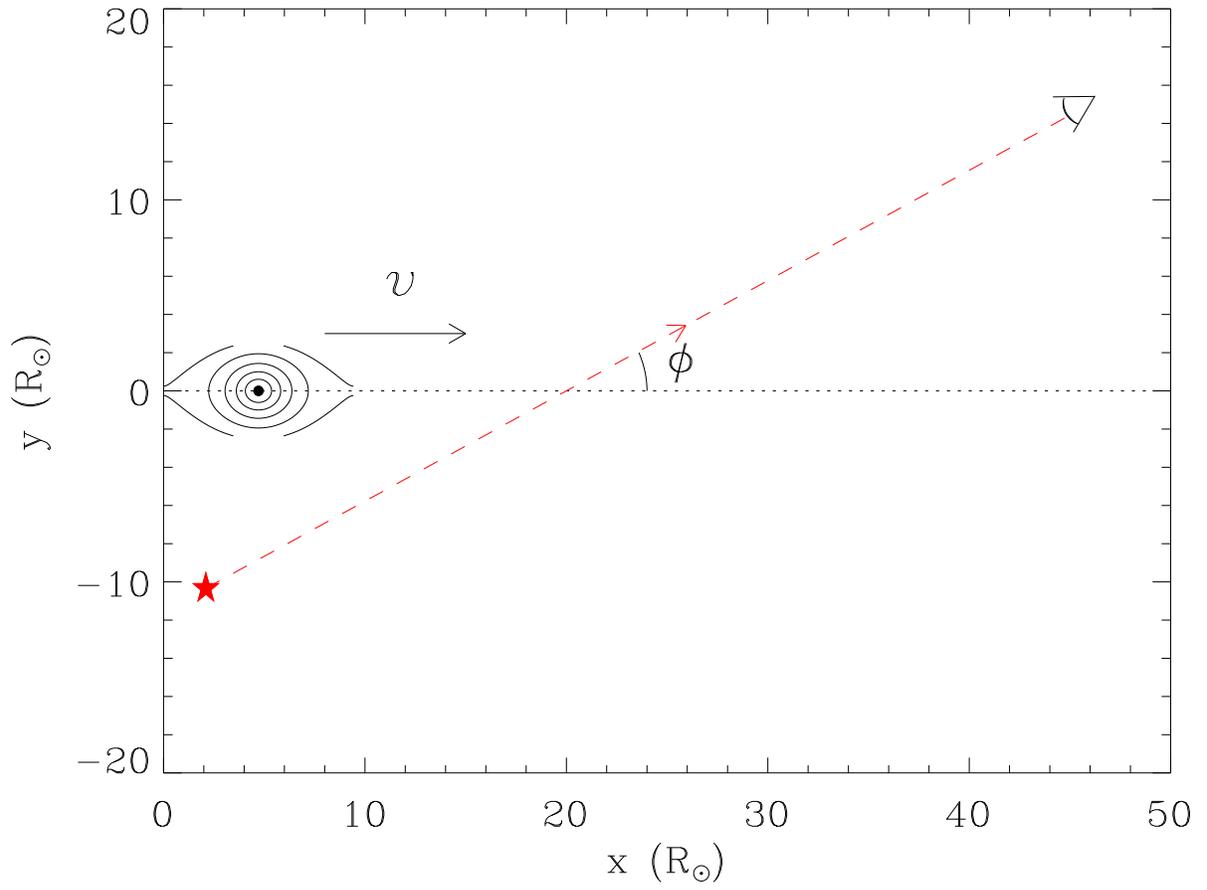} \caption{Same format as Figure~2, but for a
non-force-free flux rope embedded in a current sheet.}
\end{figure}

\clearpage

\begin{figure}
\plotone{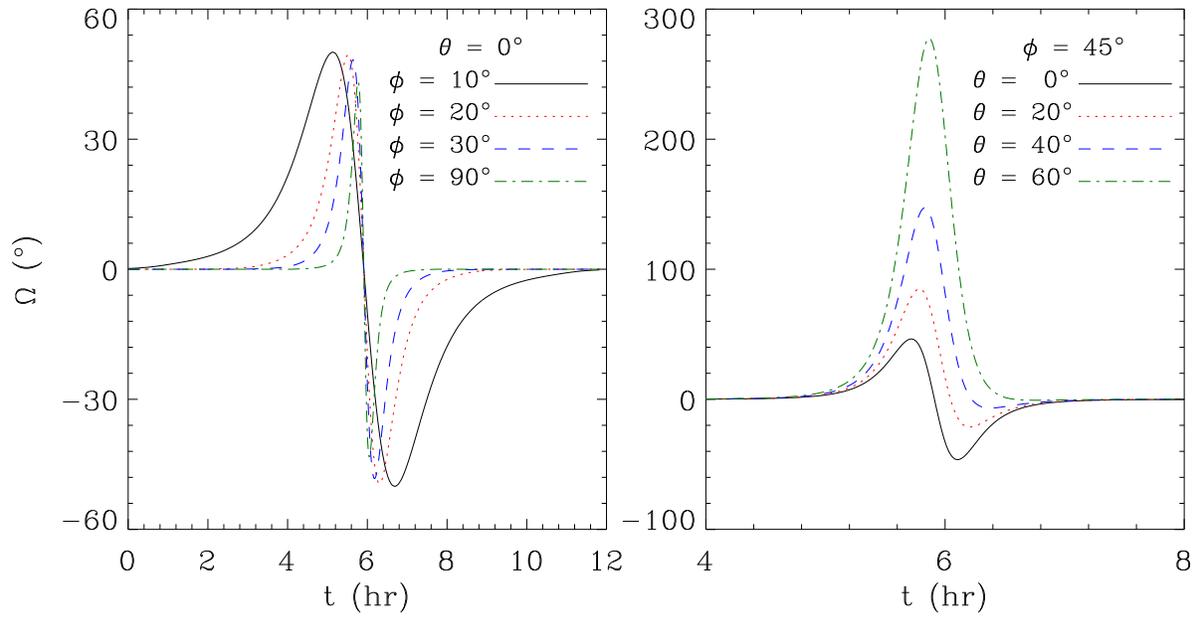} \caption{Same format as Figure~3, but for crossings
of the non-force-free flux rope.}
\end{figure}

\clearpage

\begin{figure}
\plotone{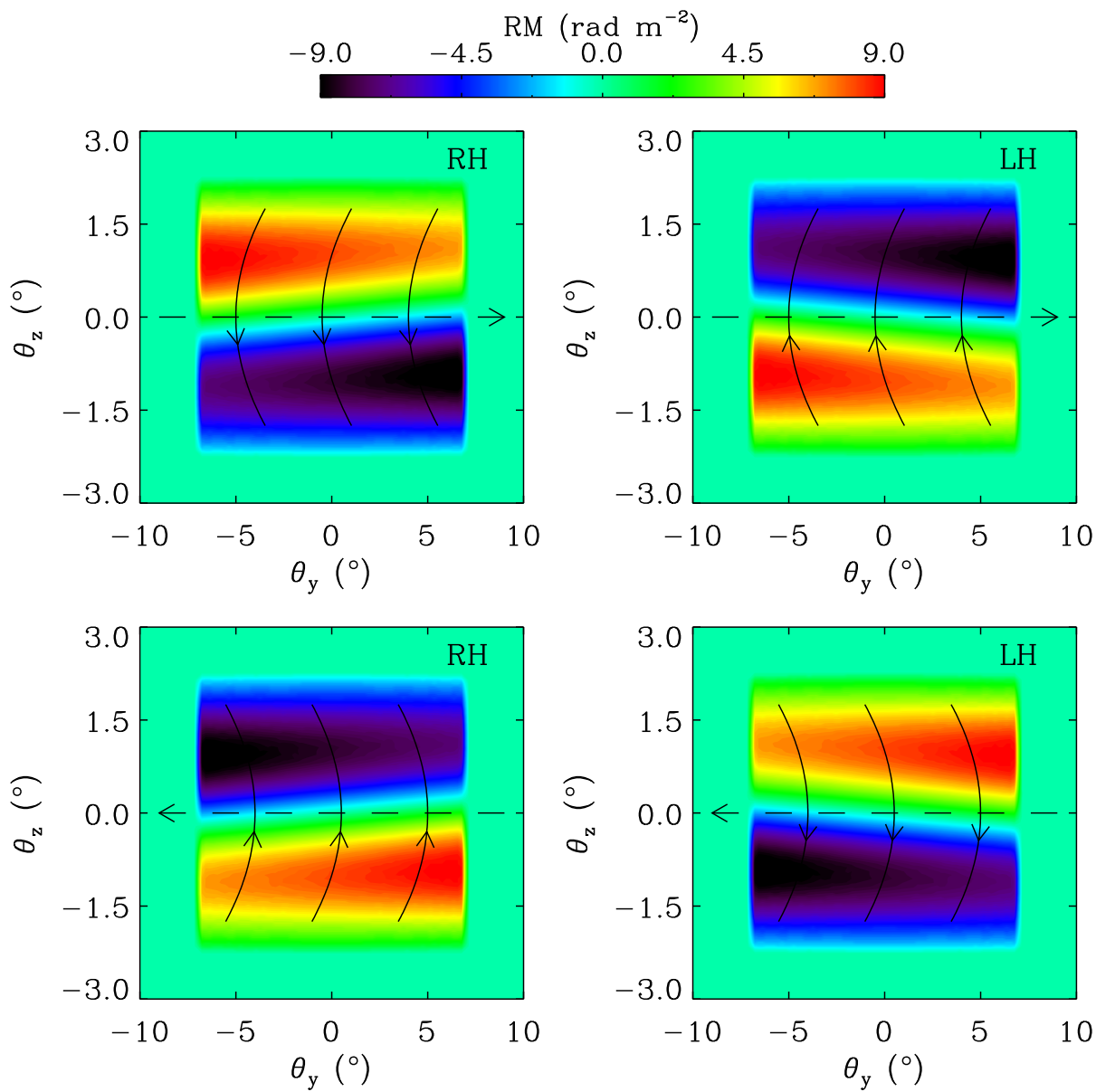} \caption{Mapping of the rotation measure
corresponding to the four configurations of a flux rope onto the sky.
The color shading indicates the value of the rotation measure. The
arrows show the directions of the azimuthal and axial magnetic
fields, from which a left-handed (LH) or right-handed (RH) helicity
is apparent. Each configuration of the flux rope has a distinct
rotation measure pattern.}
\end{figure}

\clearpage

\begin{figure}
\plotone{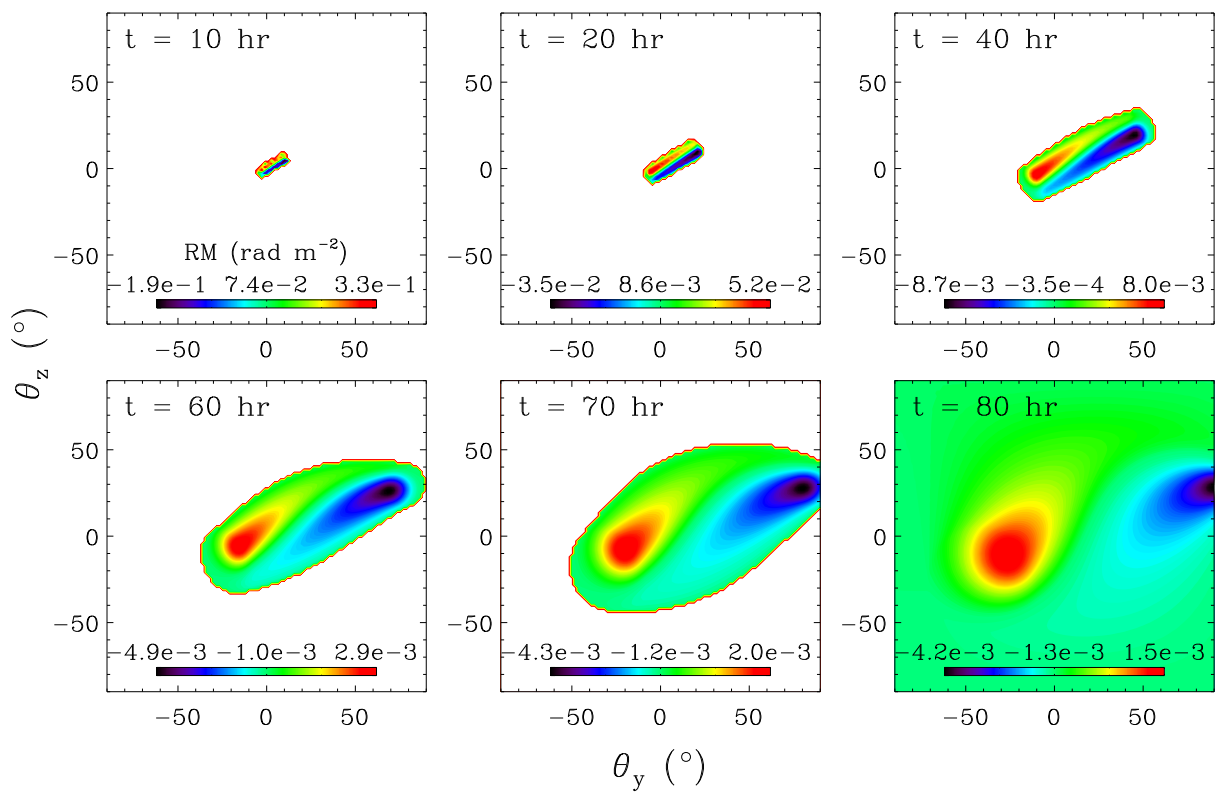} \caption{FR mapping of the whole sky at a resolution
of $\sim 3.2$ degrees as a tilted flux rope moves away from the Sun.
Note that the motion direction of the flux-rope center is not
directly toward Earth. Values of the rotation measure for each panel
are indicated by the color bar within the panel. Also shown is the
time at the top for each snapshot.}
\end{figure}

\clearpage

\begin{figure}
\plotone{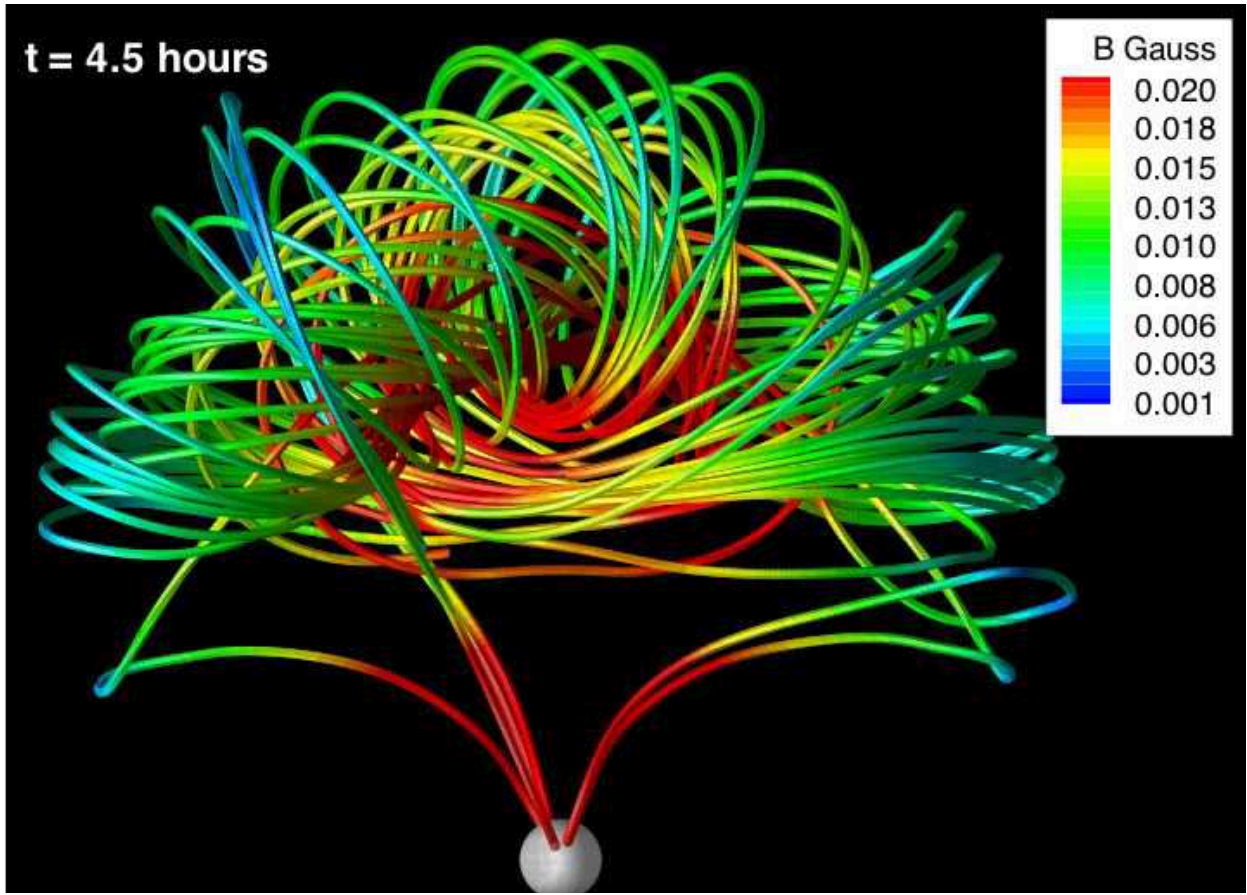} \caption{A 3-D rendering of the CME magnetic field
lines at 4.5 hours after initiation. The color shading indicates the
field magnitude and the white sphere represents the Sun.}
\end{figure}

\clearpage

\begin{figure}
\plotone{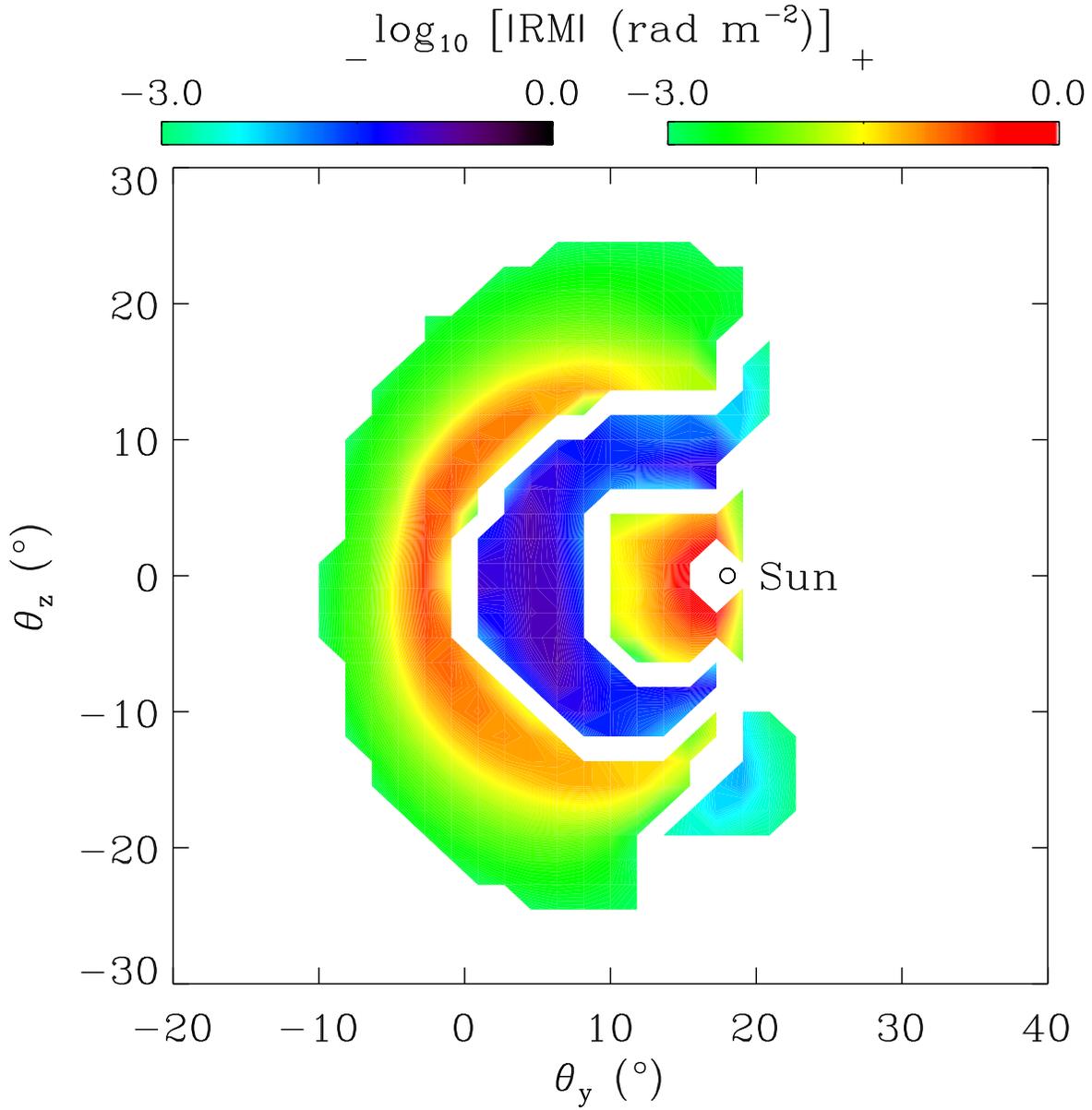} \caption{Mapping of the rotation measure difference
between the MHD simulation at 24 hours and the steady state
heliosphere. The two color bars indicate the logarithmic scale of the
absolute value of the negative (-) and positive (+) rotation measure,
respectively.}
\end{figure}

\end{document}